\documentstyle[psfig,a4,12pt]{article}
\def\mathon{\begin{equation}}
\def\mathoff{\end{equation}}
\def\arron{\begin{eqnarray}}
\def\arroff{\end{eqnarray}}
\newcommand{\p}{\partial}
\newcommand{\non}{\nonumber}
\newcommand{\ph}{\varphi}
\newcommand{\sg}{\sigma}
\newcommand{\La}{\Lambda}
\newcommand{\eps}{\epsilon}

\newcommand{\we}{\wedge}
\newcommand{\sdg}{\sqrt{g}}
\newcommand{\sdgh}{\sqrt{\hat{g}}}
\newcommand{\gh}{\hat{g}}

\begin{document}

\begin{titlepage}
\thispagestyle{empty}
\begin{flushright}
LMU-TPW 96-27\\
TUM-HEP-258/96\\
SFB-375/122\\
hep-th/9610166\\
October 1996\\
\end{flushright}
\begin{center}
\vspace*{1cm}
\renewcommand{\thefootnote}{\fnsymbol{footnote}}
{\LARGE{ \bf String Kaluza-Klein cosmologies with RR-fields}\footnote{Work 
supported by the DFG through SFB 375-95 and the European Commission TMR 
programs ERBFMX-CT96-0045 and ERBFMRX-CT96-0090}}
\vskip1cm
R. Poppe$^1$\footnote{e-mail: Rudolf.Poppe@physik.tu-muenchen.de}, 
S. Schwager$^2$\footnote{e-mail: Stefan.Schwager@physik.uni-muenchen.de}
\renewcommand{\thefootnote}{\arabic{footnote}}
\vskip.5cm
{\sl $^1$ Physik Department, Technische Universit\"at M\"unchen \\
85748 Garching, Germany}
\vskip.2cm
{\sl $^2$ Sektion Physik, Universit\"at M\"unchen \\
Theresienstr. 37, 80333 M\"unchen, Germany}
\end{center}
\vskip1cm


\begin{abstract}
We construct 4-dimensional cosmological FRW--models by compactifying a black 
5-brane solution of type IIB supergravity, which carries both magnetic 
NS-NS-charge and RR-charge. 
The influence of nontrivial RR-fields on the dynamics of the cosmological 
models is investigated.
\end{abstract}
\end{titlepage}
\newpage
\setcounter{page}{1}

\section{Introduction}
Cosmological models obtained by dimensional reduction of 10-dimensional 
supergravity theories are an active field of research 
\cite{LEA1}--\cite{KAL}. 
Since supergravity is the low energy limit of string theory, these models 
should provide a way towards understanding the quantum effects in cosmology. 
So far, there has been much activity in investigating models including fields 
of the NS-NS-sector. The discovery, that the conformal field theory of 
p-brane-like superstring backgrounds charged under RR-gauge fields is that of 
open strings with Dirichlet boundary conditions \cite{POL1,POL2}, has made the 
RR-field contents of type II string theories a very interesting subject for 
string cosmological models. 

Recently, in a first study of cosmological models including RR-fields, it was 
found that inflating phases can occur in a certain range of parameters which 
label the solutions to the equations of motion \cite{LOW}. The NS-NS-3-form 
field strength had been set to zero in the model and the compactification 
ansatz was chosen to contain a number of maximally symmetric flat subspaces.
The two branches of the solutions were separated and curvature singularities 
appeared in each branch.

In this paper we construct curved 4-dimensional cosmological models containing
a NS-NS-3-form field strength as well as nontrivial type IIB RR-fields. 
To construct the 4-dimensional model, we use a 5-brane-solution of type IIB 
supergravity whose RR-sector consists of a 3-form and a scalar. 
By SL(2,R)-invariance of the action, which is realized as a SL(2,Z)-subgroup 
in IIB superstring theory, we generate 
5-brane-solutions with nontrivial RR-fields and compactify the transverse 
coordinates on a 5-torus, giving a black hole in 5 dimensions. After 
generalizing the black hole solution to constant curvature solutions we 
compactify to 4 dimensions, obtaining isotropic FRW-metrics together with 
nontrivial RR-fields. 

The outline of the paper is as follows: In section 2 we 
introduce the
SL(2,R)-transformed 5-brane-solution of type IIB supergravity and compactify 
it on a 5-torus. In section 3 we make one further compactification 
to 4 
dimensions and obtain FRW-models including nontrivial RR-fields in this way. 
In section 4 we investigate the 
influence of the RR-fields on the cosmological models with a detailed analysis 
of the inflationary phase. Section 5 is devoted to the extremal 
limit of the model.

\section{The 5-brane-solution}
Our starting point is the bosonic part of the 10-dimensional 
SL(2,R)-invariant type IIB supergravity action \cite{SCHWAR}
\mathon \label{IIB}
S_{10}=\int d^{10}x \sqrt{G} \left\{ e^{-2\Phi}\left( R + 4 (\p\Phi)^2 -
       \frac{1}{12} H_1^2 \right) -
       \frac{1}{2} (\p\chi)^2 - \frac{1}{12} (H_1\chi + H_2)^2 \right\}.
\mathoff
In this action the string frame metric $G_{MN}$, the dilaton $\Phi$ and the
three form field strength $H_1=dB_1$ are the familiar fields contained in the 
NS-NS sector of all 10-dimensional string theories. The three form field 
strength $H_2=dB_2$ and the scalar $\chi$ belong to the RR-sector. 
The bosonic field content 
of IIB superstring theory also includes a selfdual five form field strength, 
which we have set to zero in order to be able to write a covariant action.
A 5-brane-solution to this action is \cite{GM,HS}
\mathon\label{sol10m}
ds_{(10)}^2= 
             -\frac{1-\left(\frac{r_+}{r}\right)^2}
                   {1-\left(\frac{r_-}{r}\right)^2} dt^2+     
             \frac{dr^2}{\left(1-\left(\frac{r_+}{r}\right)^2\right)
                         \left(1-\left(\frac{r_-}{r}\right)^2\right)} +
             r^2 d\Omega_3^2 +\delta_{ij}dx^idx^j
\mathoff
for the 10-dimensional line element in the string frame and
\mathon\label{sol10f}
\Phi=-\frac{1}{2}\ln\left(1-\left(r_-/r\right)^2\right),\;\;\;
H_1=Q\eps_3,\;\;\;H_2=\chi=0,
\mathoff
for the remaining background fields. The magnetic charge is $Q=2r_+r_-$ and
$d\Omega_3$ and $\eps_3$ are the line element and volume element of a 3-sphere,
respectively, i.e
\arron 
d\Omega_3&=&d\psi^2+\sin^2\psi (d\theta^2+\sin^2\theta d\ph^2) \non\\
\eps_3&=&\sin^2\psi\sin\theta d\psi\we d\theta\we d\ph.
\arroff
The RR-fields $H_2$ and $\chi$ are not
excited in this solution. We use SL(2,R)-invariance of the action
(\ref{IIB}) to generate nontrivial RR-fields, making the solution
(\ref{sol10m},\ref{sol10f}) an intrinsic IIB solution.

\subsection{The SL(2,R)-transformed 5-brane}
A manifest SL(2,R)-invariant form of (\ref{IIB}) was given in \cite{SCHWAR,BB}
and also used there to construct SL(2,R)-transformed 5-brane-solutions (see also
\cite{AAT,HMS,CH} for black hole solutions of type II stringtheory). We
do not Hodge-dualize the RR-3-form here, instead we construct a solution
possessing two magnetic charges. 

Following \cite{SCHWAR}, the action of the SL(2,R)-matrix
$\La=\left(\begin{array}{cc}a&b\\c&d\end{array}\right), ad-bc=1$ 
on the background fields is
\arron
\chi' &=&\frac{bd+ace^{-2\Phi}}{d^2+c^2e^{-2\Phi}} \non\\
e^{-\Phi'} &=& \frac{e^{-\Phi}}{d^2+c^2e^{-2\Phi}} \non\\
H_1' &=& dH_1 \non\\
H_2' &=& -bH_1.
\arroff 
This transformation clearly includes S-duality for the choice $a=d=0, -b=c=1$,
which interchanges the two 3-forms and inverts the string coupling 
$g_s=e^\Phi \to 1/g_s$.
In detail, the transformed scalar fields are 
\arron\label{SL2sol10f}
\chi'&=&\frac{bd+ac(1-r_-^2/r^2)}{d^2+c^2(1-r_-^2/r^2)}  \non\\
\Phi'&=&-\frac{1}{2}\ln\left(1-\left(r_-/r\right)^2\right)+
         \ln\left(d^2+c^2\left(1-\left(r_-/r\right)^2\right)\right)
\arroff
and the string frame line element changes to
\mathon\label{SL2sol10m}
ds'^2_{(10)}=\sqrt{d^2+c^2(1-r_-^2/r^2)}ds_{(10)}^2.
\mathoff
The Einstein frame metric is not changed by the SL(2,R) transformation.

\subsection{The 5-dimensional black hole}
We compactify this solution to five dimensions by wrapping the five
transverse coordinates of the 5-brane over a 5-torus, i.e.\ we make the
ansatz for the string frame metric
\mathon
ds^2=\gh_{\mu\nu}dx^\mu dx^\nu + e^A \delta_{ij}dx^i dx^j,
\mathoff
where $\gh_{\mu\nu}$ is a 5-dimensional Lorentz metric, $x^i$ are coordinates 
on the 5-torus and $A$ depends on the $x^\mu$ only. We neglect the moduli
coming from compactifying the three forms as well as vector fields coming from
the metric.
The effective 5-dimensional action is
\arron\label{5DString}
S_5&=&\int d^5x \sdgh\; \Bigg\{e^{-\phi}\left(\hat{R} + (\p\phi)^2 -
      \frac{1}{5}(\p\sg)^2 -
      \frac{1}{12} H_1^2\right) - \non\\
&&{}- \frac{1}{2} e^{\sg} \left( (\p\chi)^2  + 
      \frac{1}{6}(H_1\chi+H_2)^2\right)\Bigg\},
\arroff
where we have used the rescaled modulus field $\sg=\frac{5}{2}A$ and the 
5-dimensional dilaton $\phi$ is given by $\phi=2\Phi-\sg$.
From the transformed 5-brane-solution (\ref{SL2sol10f}, \ref{SL2sol10m}) we 
get a 5-dimensional solution
\arron\label{SL2sol5String}
ds_5^2&=&\sqrt{d^2+c^2\left(1-(r_-/r)^2\right)}
\Bigg[ -\frac{1-\left(\frac{r_+}{r}\right)^2}
             {1-\left(\frac{r_-}{r}\right)^2}dt^2+
        \frac{dr^2}{\left(1-\left(\frac{r_+}{r}\right)^2\right)
                    \left(1-\left(\frac{r_-}{r}\right)^2\right)}+ \non\\
&&{}+r^2 d\Omega_3^2 \Bigg] \non\\
H_1&=&dQ\eps_3,\;\;\;\;\;\;H_2=-bQ\eps_3      \non\\
\phi&=&\ln\frac{\left(d^2+c^2\left(1-(r_-/r)^2\right)\right)^{3/4}}
                         {1-(r_-/r)^2}        \non\\
\sg&=&\frac{5}{2}\ln\left(d^2 + c^2\left(1-(r_-/r)^2\right)\right) \non\\
\chi&=&\frac{bd+ac\left(1-(r_-/r)^2\right)}
            {d^2+c^2\left(1-(r_-/r)^2\right)}.
\arroff
For $\Lambda$ the identity transformation, i.e.\ vanishing RR-fields $H_2$ and 
$\chi$, this is just the 5-dimensional
black hole solution of \cite{HS}, written in the string frame.
In these equations $r_+$ and $r_-$ are the values of the parameter $r$ at the
outer and inner horizon, respectively. The metric has a curvature singularity 
at $r=0$ and also at the spacelike surface $r=r_-$ whereas
$r_+$ is a regular event horizon. This is a black hole 
solution for $r_+ \geq r_-$ only, in the
other case the solution describes a naked singularity. 
A particular interesting solution is the extremal limit at $r_+=r_-$. We will 
discuss the non-extremal and extremal cases in a cosmological context in the 
next sections.   

\section{4-dimensional FRW-cosmologies}
Following \cite{BF}, where this solution was generalized to constant curvature
spaces, we can produce a 4-dimensional cosmological solution from
this black hole by Kaluza-Klein compactification of the time coordinate and 
re-interpretation of the radius coordinate as 4-dimensional time. 
In order to get the right signature after compactification we must change
signature from $(-,+,+,+,+)$ to $(+,-,+,+,+)$ in the 5-dimensional solution.
The generalization of (\ref{SL2sol5String}) to constant curvature spaces is 
given by 
\arron\label{sol5gen}
ds_5^2&=&\sqrt{d^2+c^2\left(1-(t_-/t)^2\right)}
\Bigg[ \frac{-k+\left(\frac{t_+}{t}\right)^2}
            {1-\left(\frac{t_-}{t}\right)^2}dy^2-
        \frac{dt^2}{\left(-k+\left(\frac{t_+}{t}\right)^2\right)
                    \left(1-\left(\frac{t_-}{t}\right)^2\right)}+ \non\\
&&{}+t^2 d\Omega_k^2 \Bigg] \non\\
H_1&=&dQ\eps_k,\;\;\;\;\;\;H_2=-bQ\eps_k,\;\;\;\;\;Q=2t_+t_-  \non\\
\phi&=&\ln\frac{\left(d^2+c^2\left(1-(t_-/t)^2\right)\right)^{3/4}}
                         {1-(t_-/t)^2}        \non\\
\sg&=&\frac{5}{2}\ln\left(d^2 + c^2\left(1-(t_-/t)^2\right)\right) \non\\
\chi&=&\frac{bd+ac\left(1-(t_-/t)^2\right)}
            {d^2+c^2\left(1-(t_-/t)^2\right)},
\arroff
where $k\in\{-1,0,1\}$ labels the constant curvature three-spaces with line 
element and volume element
\arron
d\Omega_k&=&d\psi^2+\left(\frac{\sin\sqrt{k}\psi}{\sqrt{k}}\right)^2
            (d\theta^2+\sin^2\theta d\ph^2), \non\\
\eps_k&=&\left(\frac{\sin\sqrt{k}\psi}{\sqrt{k}}\right)^2\sin\theta
   d\psi\we d\theta\we d\ph.
\arroff
It should be emphasized that the time coordinate in this solution is the radius
coordinate of the 5-dimensional black hole and the signature is different from
usual black hole physics.

We reduce to 4 dimensions using the ansatz 
\mathon
ds_5^2=Y^2dy^2+ds_4^2
\mathoff
where the $S^1$-modulus $Y^2$ turns out to be
\mathon
Y^2(t) = \frac{-k+(t_+/t)^2}{1-(t_-/t)^2}
         \sqrt{d^2+c^2\left(1-t_-^2/t^2\right)}.
\mathoff
By introducing $\rho=\ln Y$ we arrive at the 4-dimensional action
\arron \label{CosString}
S_4&=&\int d^4x\sdg\;\Bigg\{ e^{-2\ph}\left(R+4(\p\ph)^2-(\p\rho)^2 -
       \frac{1}{5}(d\sg)^2 -\frac{1}{12}H_1^2\right) - \non\\
&&{}-\frac{1}{2}e^{\rho+\sg}\left((\p\chi)^2+
     \frac{1}{6}(H_1\chi+H_2)^2\right) \Bigg\}.
\arroff
The 4-dimensional dilaton $\ph$ is given by $\ph=\frac{1}{2}(\phi-\rho)$.
The 5-dimensional black hole (\ref{sol5gen}) reduces to the 4-dimensional 
FRW-solution 
\arron\label{FRWString}
ds_4^2&=&\sqrt{d^2+c^2\left(1-t_-^2/t^2\right)}
\left(
-\frac{dt^2}{\left(-k+\left(\frac{t_+}{t}\right)^2\right)
             \left(1-\left(\frac{t_-}{t}\right)^2\right)}+
      t^2 d\Omega_k^2
\right)          \non\\
H_1&=&dQ\eps_k,\;\;\;H_2=-bQ\eps_k,\;\;\;Q=2t_+t_- \non\\
\rho&=&\frac{1}{2}\ln\left(
       \frac{\left(-k+\left(\frac{t_+}{t}\right)^2\right)
             \sqrt{d^2+c^2\left(1-t_-^2/t^2\right)}}
            {1-\left(\frac{t_-}{t}\right)^2}
       \right)      \non\\
\ph&=&\frac{1}{4}\ln\left(
      \frac{d^2+c^2\left(1-t_-^2/t^2\right)}
           {\left(-k+\left(\frac{t_+}{t}\right)^2\right)
      \left(1-\left(\frac{t_-}{t}\right)^2\right)}
      \right)\non\\
\sg&=&\frac{5}{4}\ln\left(d^2+c^2\left(1-t_-^2/t^2\right)\right) \non\\
\chi&=&\frac{bd+ac\left(1-(t_-/t)^2\right)}
            {d^2+c^2\left(1-t_-^2/t^2\right)}
\arroff
For the special choice of SL(2,R)-parameters $a=d=1,b=c=0$, the RR-fields $H_2$ 
and $\chi$
and the $T^5$-modulus field $\sg$ are zero and we recover the cosmological 
models described in \cite{BF}. These models have been studied in detail there,
both in the string and Einstein frame. Since the SL(2,R)-transformation does
not affect the Einstein metric, we concentrate on the string frame in the 
following. Let us briefly recall the properties of the model in the pure 
NS-NS-case, i.e.\ without contributions from the RR-fields.
The solution describes a closed $(k=+1)$,
open $(k=-1)$, or flat $(k=0)$ cosmology respectively. In the string frame, 
for $k=+1$, the three space oscillates\footnote{This
can be seen after changing to conformal time by the transformation 
(\ref{conftime}).} between the extrema $t_\pm$. 
For $k=0,-1$ the solution is not oscillating, the extension of the corresponding 
three spaces is bounded from below at $t=t_-$ and tends to infinity as 
$t\to\pm\infty$. The dilaton $\ph$ is always divergent at $t=t_-$, corresponding 
to the minimal extension of the three spaces. For $k=+1$, the dilaton is also
divergent at $t=t_+$, while for $k=0,-1$, where the time region has no upper 
bound, it tends to zero for $t\to\pm\infty$. The $S^1$-modulus $\rho$ is
divergent at $t=t_-$, indicating decompactification to 5 dimensions at
this points. For $k=+1$, the $S^1$-radius becomes zero at the maximum extension
of three-space. In the other cases $k=0,-1$, the modulus tends to a constant
value as $t\to \pm\infty$. 

In the Einstein frame, the curvature scalar is singular at $t=t_-$ in all 
three cases and  also at $t=t_+$ in the case $k=+1$. The extension of 
three-space shrinks to zero at these points, indicating a big-crunch or 
big-bang.

\section{Effects of the RR-fields on the cosmological models}
Let us now study the qualitative changes caused by adding nontrivial RR-fields
to the solution described above. As can be seen from the dilaton expression in 
(\ref{FRWString}),
in the case $d=0$ one divergence cancels leaving the 
expression for the dilaton
\mathon
\ph_{d=0}=\frac{1}{4}\ln\left(\frac{c^2}
          {-k+\left(\frac{t_+}{t}\right)^2}\right).
\mathoff  
For $k=+1$, there is still a divergence at $t=t_+$, but the divergence at
$t=t_-$ has disappeared in all cases $k=-1,0,+1$. The case $d=0$ corresponds to
zero NS-NS-charge and a constant RR-scalar $\chi=-ab$. We want to see how the 
scale factors of the FRW-models behave
in this special case. For this we change to a coordinate system which is 
usually used to study FRW-cosmologies
\mathon
ds^2=-d\tau^2+a^2(\tau)d\Omega_k^2.
\mathoff
The comoving time coordinate $\tau$ is defined by a solution to the
differential equation
\mathon\label{tau}
d\tau^2=\frac{\sqrt{d^2+c^2\left(1-t_-^2/t^2\right)}}
             {\left(-k+\left(\frac{t_+}{t}\right)^2\right)
              \left(1-\left(\frac{t_-}{t}\right)^2\right)}dt^2,\;\;\;\;
              t(\tau=0)=t_-
\mathoff
and the scale factor $a(\tau)$ is 
\mathon\label{scalefactor}
a^2(\tau)=t^2(\tau)\sqrt{d^2+c^2\left(1-t_-^2/t^2(\tau)\right)}.
\mathoff
For $d=0$ the scale factor $a(\tau)$ goes to zero at $\tau=0$.
Calculating $\dot{a}(\tau)$ we find the two different behaviors 
\mathon
\lim_{\tau\to 0}\frac{da(\tau)}{d\tau}=\left\{\begin{array}{r@{\quad:\quad}l}
                                           0      & d\not= 0 \\
                                           \infty & d=0
                                       \end{array}\right.
\mathoff
So by switching off the NS-NS-3-form $H_1$ completely, we produce a 
singularity in the string 
frame scale factor, whereas a finite value of the NS-NS-charge smoothes out 
this singularity. 
At the same time we find that the $T^5$-compactification radius is zero for
$d=0$ at $t=t_-$, whereas the $S^1$-radius is still divergent there.
The dilaton $\ph$ takes the minimum value at $t=t_-$ for $d=0$ and it
diverges there for $d\not= 0$. The magnetic charge of the NS-NS-3-form prevents 
the universe from collapsing, as was already recognized in \cite{KB}. 
Without NS-NS-charge, the scale factor exhibits a singularity regardless of the
existence of RR-charge.

The qualitative behavior of the dilaton-singularities does not 
change when RR-fields are added to a non-zero NS-NS-field strength. 
The minimum extension of the universe at $t=t_-$ is $t_-\sqrt{d}$ and for 
$k=+1$ the maximum extension is given by 
$t_-\left(d^2+c^2(1-t_-^2/t_+^2\right)^{1/4}$.
However, the duration and effectivity of a possible initial inflationary phase 
depend on the parameters 
c and d as well as on $t_-$ and $t_+$, as can be seen from the equations 
(\ref{tau}) and (\ref{scalefactor}).
We will discuss this dependence in detail now.

\subsection{The inflationary phase}
In the presence of a NS-NS-charge the string frame scale factor of the 
cosmology is free of singularities, although in the Einstein frame the
familiar curvature singularity close to the time $t=t_{-}$ appears. 
It is interesting to investigate how this feature is affected by adding 
RR-fields and whether a phenomenologically interesting inflationary phase can 
occur.
From a phenomenological point of view one can argue that the Einstein 
frame is the preferred coordinate system. On the other hand one can
treat both frames on equal footing, as has been done in recent years
in numerous analyses where the transition of a superinflationary 
pre-big bang phase into a a FRW type cosmology
was worked out in the string frame 
\cite{GREX1}--\cite{GREX7}.

In the following we will examine in more detail the effect of RR-fields on the 
string frame scale factor close to the time $t=t_{-}$,
where the influence of these fields are most apparent.
It was pointed out in \cite{BF} that in the asymptotic limit 
$t\rightarrow t_{-}$
the expansion scales like $a(\tau) \sim \tau^{2} + const.$ and from the
discussion in the previous section we know that the S-dual
model contains a singularity in the scale factor at this point. 
One can easily imagine that close to $\tau=0$
a short phase of superluminal expansion might occur in the sense of power-law
inflation with $a(\tau) \sim \tau^p$ 
and $p$ sufficiently large to ensure an expansion that could 
be of phenomenological interest.

As a measure to describe an accelerated expansion of the universe we
use the quantity $\zeta\equiv\ddot{a}/a=\dot{H}\,+H^2 =
\frac{p(p-1)}{\tau^2}$. Here $H$ is the Hubble parameter
$H=\dot{a}/a$, a dot means the derivative with respect to
$\tau$ and the last part of the equation holds if the expansion is of
power law type, $a(\tau) \sim \tau^p$. In the general case 
we are not able to find analytical expressions for the quantity
$\zeta$, because equation (\ref{tau}) can not be integrated
directly.  
Instead we will give numerical solutions for a certain
choice of parameters in fig.\ 1. It is shown how the acceleration
changes with transformations that lie within a SO(2,R)-subgroup
of SL(2,R), parametrized by the rotation angle $\beta$. An increasing rotation 
angle corresponds to an
enlargement of the fields of the RR-sector and simultaneously
a reduction of the NS-NS-3-form field strength. 
We observe that the initial acceleration is very much enhanced by exchanging
NS-NS-charge by RR-fields but a subsequent decelerating phase becomes also 
gradually more important and thus taking off much of what was gained in the 
accelerating phase.
At the S-dual point we encounter again
the transition from a non-singular to the singular universe and
$\zeta\rightarrow -\infty $. The number of e-foldings ${\cal N}$
produced during the expansion increases only slightly by enlarging
the RR-fields and remains of the order of one for almost
all values of $\beta$.
Since ${\cal N}$ must exceed a certain limit as a prerequisite for a 
viable inflationary scenario, none of the cosmologies fulfills this
requirement.

\begin{figure}[hbt]
\centerline{\psfig{figure=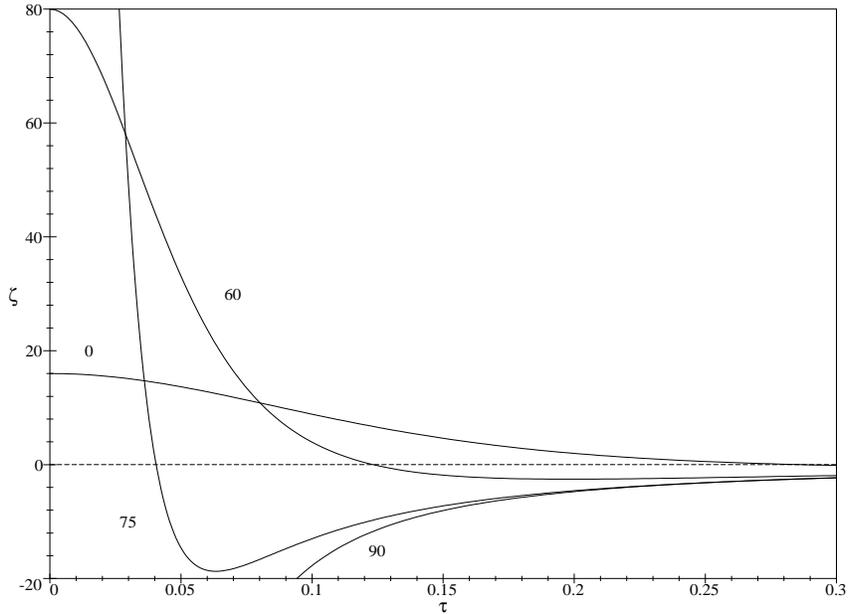,height=100mm,width=80mm}}
\caption{$\zeta\equiv\ddot{a}/a=\dot{H}\, + H^{2}$ for several SO(2)-rotations.
         (Rotation angles in degrees, 0 means zero RR-charge, 90 means zero
         NS-NS charge.) 
         The spatial section was assumed to be flat ($k=0$) and the 
         time parameters were chosen to be $t_{-}=1$ and $t_{+}=4$.}
\end{figure}

An expression for the number of e-foldings produced
between the time $t_{-}$ and $t$ can easily be derived from (\ref{scalefactor})
\mathon
{\cal N} (t;c)=\ln\,\left(\frac{t}{t_{-}}\right) +\frac{1}{4}\,\ln\left(
d+\frac{c^{2}}{d}\,
\left( 1-\left(\frac{t_{-}}{t}\right)^{2} \right)\right).
\mathoff
Since the string frame metric is only influenced by the SL(2,R)-parameters $c$ 
and $d$, we study the case where for a fixed value of d (fixed NS-NS-charge) 
the parameter $c$ is increased.
For the specific choice $a=d=1,b=0$ and $c$ a free parameter,
it becomes apparent that the RR-fields enhances
the number of e-folds only logarithmically. However, for $t_-$ being very 
close to zero, ${\cal N}$ can become sufficiently large for having
cosmological significance.  
Upon expanding $a(\tau)$ to second order in $\tau$ it can also be seen
how the singular case is recovered, 
\mathon
a(\tau )=t_-\sqrt{d} - \frac{1}{2 t_-\sqrt{d}}
\left(k-\left(\frac{t_{+}}{t_{-}} \right)^{2} \right)\,\,\left(
1+\frac{1}{2}\,\frac{c^{2}}{d^{2}} \right)\,\tau^{2} +{\cal O} (\tau^{4}).
\mathoff
The pre-factor of the quadratic term gets large for small $t_-$ and $d$ and
for large $c$. 
Also, the behavior of $\zeta$ with this choice of parameters is 
qualitatively the same as shown in fig.\ 1 for the SO(2) transformed models,
as well as for all other sets that have been investigated.  
Moreover, curved spatial sections ($k=\pm 1$) have only little influence on
the qualitative picture. 

\section{Extremal limit}
It was shown in \cite{BF} that for vanishing RR-fields the 5-dimensional 
solution (\ref{sol5gen}) has 
an extremal limit $t_+=kt_-$, where the theory corresponds to a WZW 
model\footnote{In this limit the theory decouples in a 
direct product of a 3d (spherical) part and a 2d ($\eta, y$) part. 
For $k=+1,-1$, the 2d part corresponds to a $SL(2,R)/U(1)$ coset model and for
$k=0$ it is trivial.} 
and is therefore exact to all orders in $\alpha'$. Whereas for p-brane 
superstring backgrounds charged under RR gauge 
fields the corresponding conformal field theory is known to be that of open 
strings with Dirichlet boundary
conditions \cite{POL1,POL2}, the higher order contributions to the tree level 
equations of motions for non-p-brane like backgrounds still lack an explicit 
sigma model description \cite{SENG}. For this the 
extremal limit described above may not remain exact after performing an SL(2,R) 
transformation.
The corresponding 4d solution, valid for arbitrary $k$, is 
\arron
ds^2&=&t_-^2\sqrt{d^2+c^2\left(
       \frac{\sin\sqrt{k}\eta}{\sqrt{k}\;t_-}\right)^2}
       \left(-d\eta^2+d\Omega^2_k\right)\non\\
\rho&=&\frac{1}{2}\ln\left(
       \sqrt{d^2+c^2\left(
       \frac{\sin\sqrt{k}\eta}{\sqrt{k}\;t_-}\right)^2}
       \left(\frac{\sqrt{k}}{\tan\sqrt{k}\eta}\right)^2
       \right) \non\\
\ph&=&-\frac{1}{2}\ln\left(
       \frac{\cos\sqrt{k}\eta\sin\sqrt{k}\eta}
            {\sqrt{k}t_-^2\sqrt{d^2+c^2
            \left(\frac{\sin\sqrt{k}\eta}{\sqrt{k}t_-}\right)^2}}
            \right)\non\\
\sg&=&\frac{5}{4}\ln\left(d^2+c^2\left(\frac{\sin\sqrt{k}\eta}{\sqrt{k}\;t_-}
       \right)^2\right)\non\\
\chi&=&\frac{bd+ac\left(\frac{\sin\sqrt{k}\eta}{\sqrt{k}\;t_-}\right)^2}
            {d^2+c^2\left(\frac{\sin\sqrt{k}\eta}{\sqrt{k}\;t_-}\right)^2}\non\\
H_1&=&2d \sqrt{\pm k}\;t_-^2\eps_k, \;\;\;\; 
H_2=-2b \sqrt{\pm k}\;t_-^2\eps_k                  
\arroff
In this solution, $\eta$ is the conformal time defined by the transformation
\mathon\label{conftime}
t^2=t^2_-+(t_+^2-kt_-^2)\left(\frac{\sin\sqrt{k}\eta}{\sqrt{k}}\right)^2
\mathoff
and a possible constant part of the 5-dimensional dilaton in (\ref{sol5gen}) has 
to be adjusted \cite{BF}
\mathon
\phi_0 \to \phi_0 + \frac{1}{2} \ln (t_+^2-kt_-^2)
\mathoff  
in order to be able to perform the extremal limit.
For vanishing RR-fields, we recover again the solution given in 
\cite{BF}. In this limit,
the geometry is simply $R\times S^3_k$ for all three cases (see figure 2a), 
where $S^3_k$ stands for
a sphere, pseudo-sphere or flat space respectively. Adding RR-fields to the 
solution, the universe behaves different for different values of $k$. For 
$k=1$, the conformal factor is oscillating between $|d|$ and 
$\sqrt{d^2+c^2/t_-^2}$. In the other cases the universe approaches the minimum
size $|d|$ for $\eta=0$ and then expands for ever. Especially for the 
S-dual case, 
where the NS-NS-field $H_1$ is switched off, the corresponding solutions 
exhibit again singularities (see figures 2b-d). 
\begin{figure}[htb]
\begin{minipage}{8cm}
\centerline{2a}
\psfig{figure=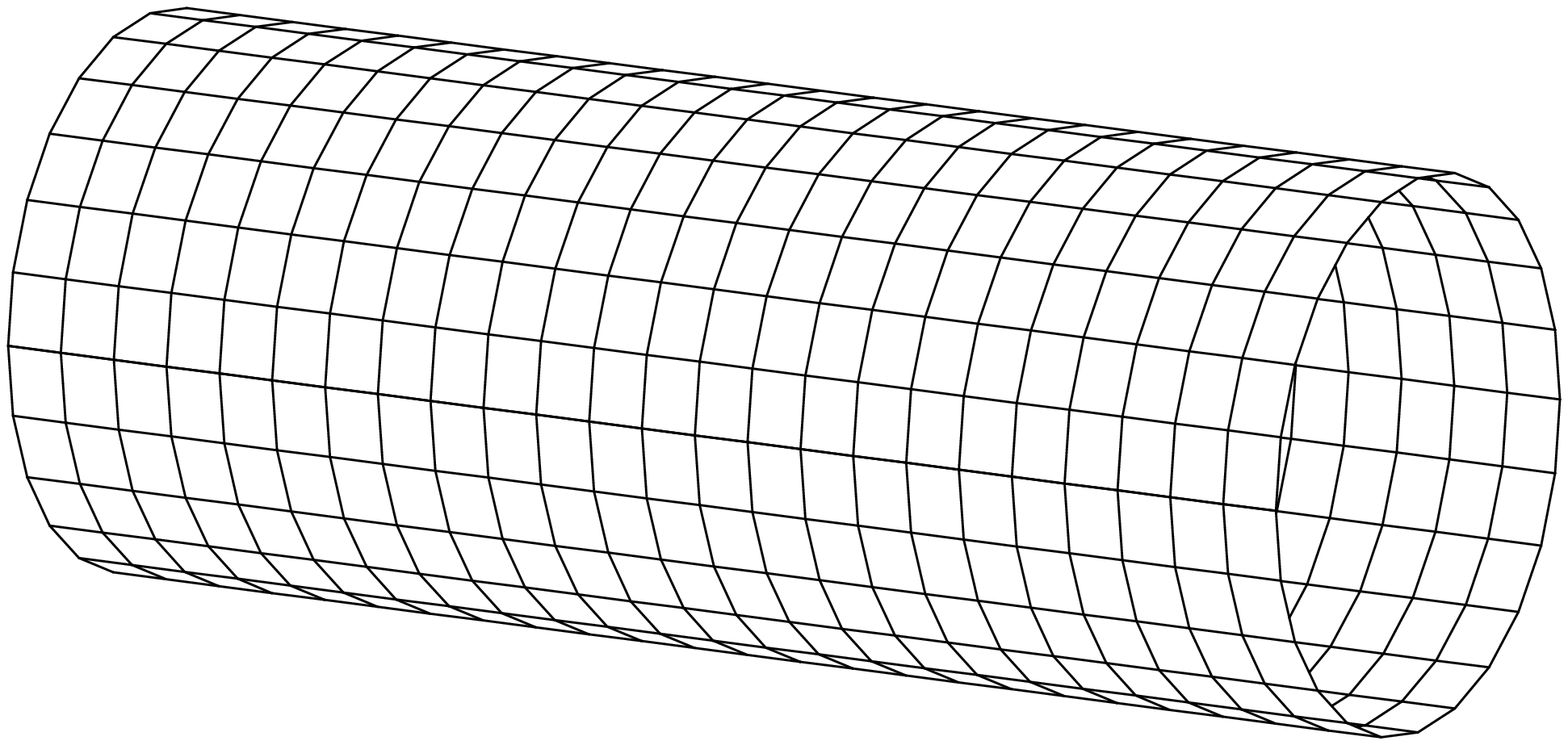,width=8cm,angle=270}
\end{minipage}\hfill
\begin{minipage}{8cm}
\centerline{2b}
\psfig{figure=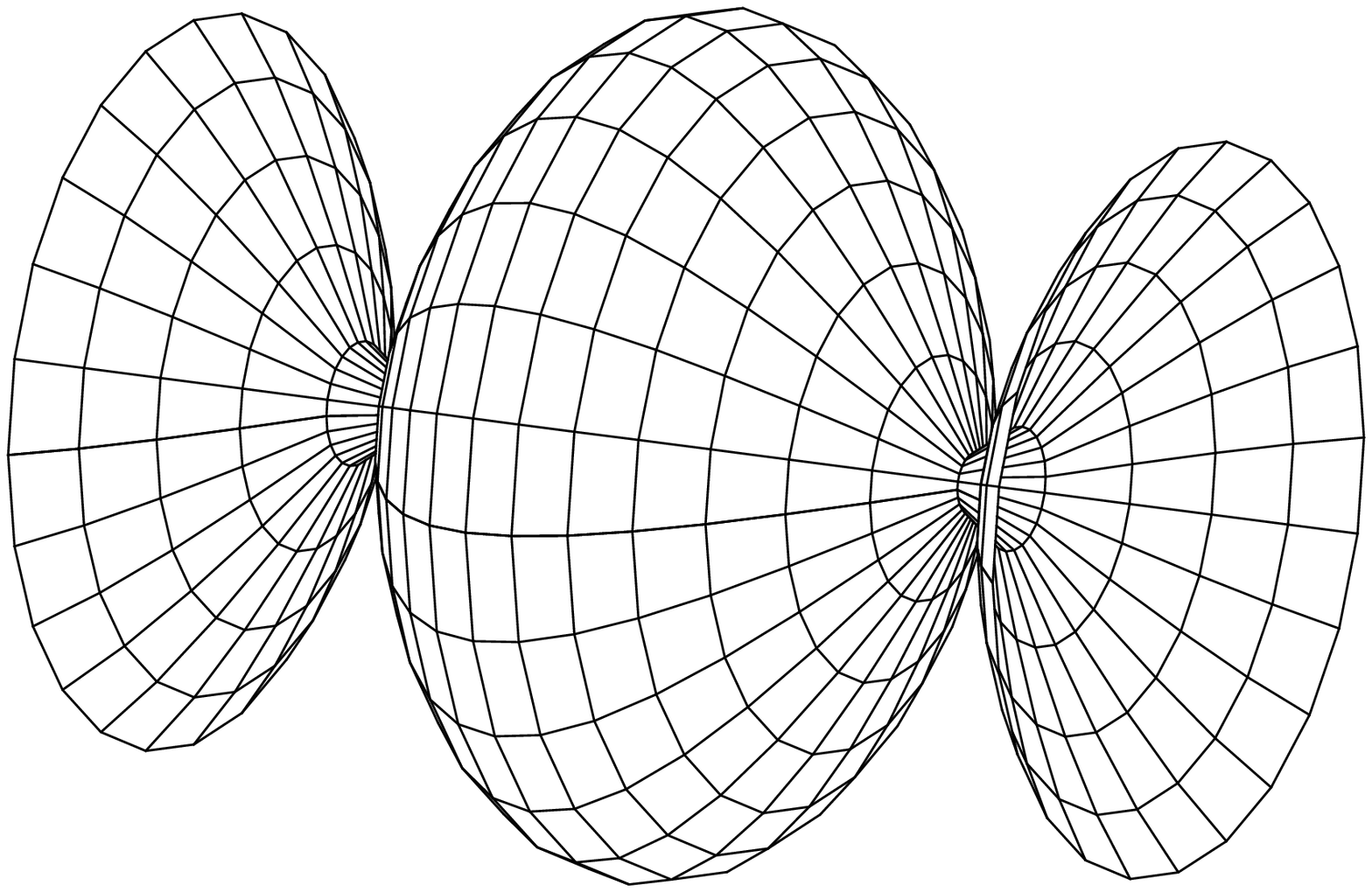,width=8cm,angle=270}
\end{minipage}\\
\begin{minipage}{8cm}
\centerline{2c}
\psfig{figure=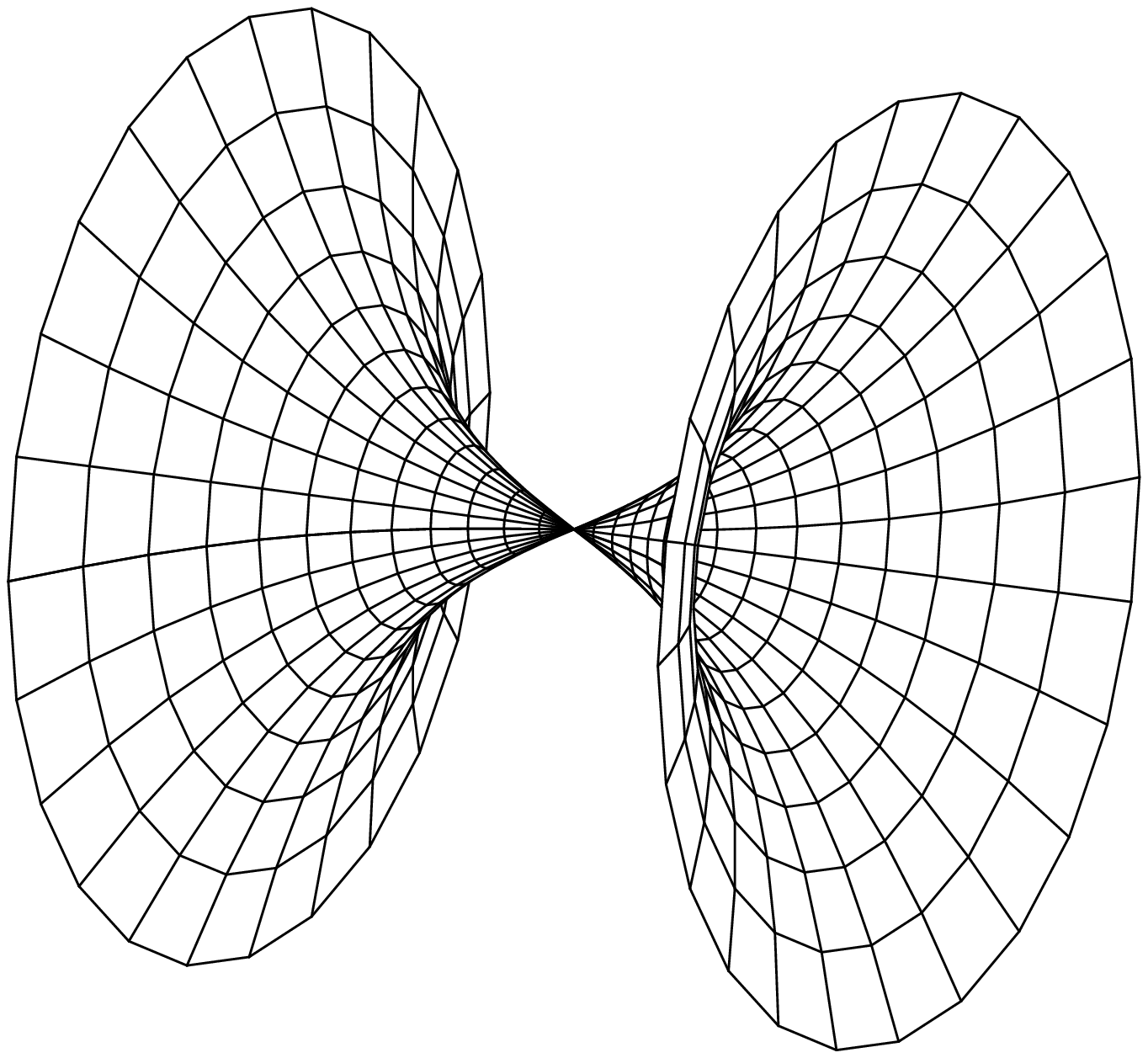,width=8cm,angle=270}
\end{minipage}\hfill
\begin{minipage}{8cm}
\centerline{2d}
\psfig{figure=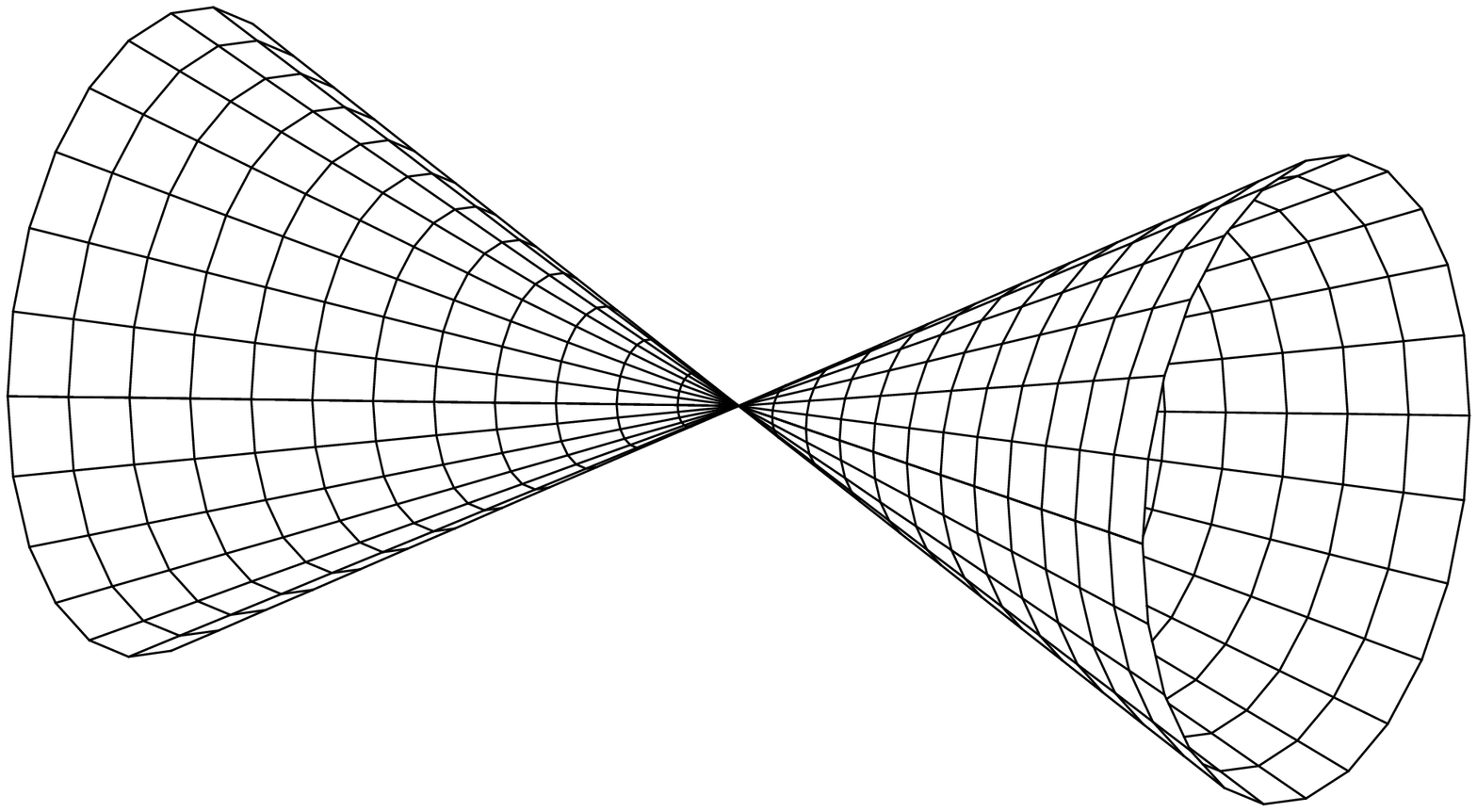,width=8cm,angle=270}
\end{minipage}
\caption{The extremal limit for pure NS-NS-charge (2a) and the extremal limits 
         for pure RR-charge for $k=+1$ (2b), $k=-1$ (2c) and $k=0$ (2d).}
\end{figure}

\section{Conclusions}
We constructed 4-dimensional FRW-cosmologies from 5-dimensional black hole 
solutions following \cite{BF}. Our models contain nontrivial RR-fields, a 
3-form and a scalar, which are 
generated by SL(2,R) transformations of a black 5-brane solution of
type IIB supergravity.

The effect of this RR-fields on the resulting cosmologies have been 
studied. For the S-dual model, where the NS-NS-charge is replaced by the
corresponding RR-charge, the dilaton singularity at the minimum extension of 
spatial 3-space disappears, but the model still decompactifies to 5 dimensions 
there. Also, other than the NS-NS-field strength, the RR-field strength cannot
prevent the universe from collapsing, and we find a singularity in the 
string frame scale factor, which is absent when the NS-NS-charge is different 
from zero.

It was further shown that the influence of RR-fields on the cosmic expansion
results in a large increase of the initial accelerating phase. It is however
followed by a subsequent deceleration phase that increases in a similar way. 
For that reason the number of e-folds 
produced during the combined phases increases only slightly, almost 
logarithmically, 
with the RR-fields being turned on. Thus, such models do not seem to be
suitable for an explanation of spatial homogeneity and isotropy, especially 
that of the CMB radiation, based on inflationary cosmology. However, for a
small fraction of parameter space, the superluminal expansion of the
physical space is effective enough to solve some of the
problems inflationary cosmology was invented for. 

We examined also the extremal limit of the model, for which the 5-dimensional
black hole is on the verge of becoming a naked singularity. Since we generalized
the 5-dimensional solution to a constant curvature solution, it is not clear 
whether the successful description of extremal black holes as being composed of
non-interacting branes, anti-branes and momentum \cite{HS2,SV} can be applied 
to this extremal limit. The effect of RR-fields added to the extremal limit
of the pure NS-NS case is to render dynamics to the steady universe. For the
S-dual case we find again singularities in the string frame scale factor.   

During writing this paper \cite{LUXU} appeared, where also cosmological 
models from toroidally compactification of type II theories are discussed. 

\section{Acknowledgment}
We would like to thank S.F\"orste and R.Dick for useful discussions.
This work was supported by the "Sonderforschungsbereich 375-95 f\"ur 
Astro-Teilchenphysik der Deutschen Forschungsgemeinschaft" and the European 
Commission TMR programs ERBFMRX-CT96-0045 and ERBFMRX-CT96-0090.
The work of S.S.\ is supported by Deutsche Forschungsgemeinschaft.


\newpage
\begin{thebibliography}{99}
\bibitem{LEA1} A.A.Tseytlin, Int. J. Mod. Phys. D 1 (1992) 223             
\bibitem{LEA2} I.Antoniadis, J.Rizos, K.Tamvakis, Nucl. Phys. B 415 (1994) 497
\bibitem{LEA3} E.J.Copeland, A.Lahiri, D.Wands, Phys. Rev. D 51 (1995) 1569
\bibitem{LEA4} R.Easther, K.Maeda, hep-th/9605173, 
               to be published in Phys. Rev. D
\bibitem{GREX1} M.Gasperini, G.Veneziano, Astropart. Phys. 1 (1993) 317
\bibitem{GREX2} M.Gasperini, G.Veneziano, Mod. Phys. Lett. A(1993) 3701
\bibitem{GREX3} M.Gasperini, G.Veneziano, Phys. Rev. D 50 (1994) 2519
\bibitem{GREX4} R.Brustein, G.Veneziano, Phys. Lett. B 329 (1994) 429
\bibitem{GREX5} N.Kaloper, R.Madden, K.A.Olive, Nucl. Phys. B 452 (1995) 677
\bibitem{GREX6} N.Kaloper, R.Madden, K.A.Olive, Phys. Lett. B 371 (1996) 34
\bibitem{GREX7} R.Easther, K.Maeda, D.Wands, Phys. Rev. D 53 (1996) 4247
\bibitem{KAL} N.Kaloper, MCGILL-96-34, hep-th/9609087 
\bibitem{POL1} J.Polchinski, Phys. Rev. Lett. 75 (1995) 4724
\bibitem{POL2} J.Polchinsky, S.Chaudhuri, C.V.Johnson, NSF-ITP-96-003, 
               hep-th/9602052
\bibitem{LOW} A.Lukas, B.Ovrut, D.Waldram, UPR-711T, hep-th/9608195
\bibitem{SCHWAR} J.H.Schwarz, Phys. Lett. B 360 (1995) 13, 
                 erratum-ibid. B364 (1995) 252 
\bibitem{GM} G.W.Gibbons and K.Maeda, Nucl.Phys. B 289 (1988) 741
\bibitem{HS} G.T.Horowitz, A.Strominger, Nucl. Phys B 360 (1991) 197
\bibitem{BB} E.Bergshoeff, J.Boonstra, Phys. Rev. D 53 (1996) 7206 
\bibitem{AAT} A.A.Tseytlin, Mod. Phys. Lett. A11 (1996) 689 
\bibitem{HMS} G.T.Horowitz, J.M.Maldacena, A.Strominger, 
              Phys. Lett. B 383 (1996) 151
\bibitem{CH} M. Cveti\v{c}, C.M.Hull,  DAMTP-R-96-31, hep-th/9606193,\\
             M. Cveti\v{c}, D. Youm, Nucl. Phys. B 476 (1996) 118
\bibitem{BF} K. Behrndt and S. F\"orste, Nucl. Phys. B 430 (1994) 441
\bibitem{KB} K.Behrndt, SLAC-PUB-6640, hep-th/9408137 
\bibitem{SENG} G.Sengupta, IMSC-96-09-27, hep-th/9609152
\bibitem{HS2} G.T.Horowitz, S.Strominger, Phys. Rev. Lett. 77 (1996) 2368
\bibitem{SV} A.Strominger, C.Vafa, Phys. Lett. B 379 (1996) 99
\bibitem{LUXU} H.L\"u, S.Mukherji, C.N.Pope, K.W.Xu, CTP-TAMU-51-96, 
               hep-th/9610107
\end{thebibliography}
\end{document}